\documentclass[conference]{IEEEtran}
\usepackage[none]{hyphenat}
\usepackage{times,helvet,courier}
\usepackage{graphics}
\usepackage{amsmath}
\usepackage{mathtools}
\usepackage{color}
\usepackage{cite}
\usepackage{breqn}
\usepackage{flushend}
\usepackage{amssymb}
\usepackage{subcaption}
\usepackage{graphicx}
\usepackage{caption}
\usepackage{amsfonts}
\usepackage[printonlyused,nolist]{acronym}
\usepackage{bm}
\usepackage{array}
\usepackage{hhline}

\makeatletter
    \newcommand{\figcaption}{\def\@captype{figure}\caption}
    \newcommand{\tabcaption}{\def\@captype{table}\caption}
\makeatother

\newcommand{\mat}[1]{\ensuremath{{\mathbf{#1}}}}
\newcommand{\matc}[1]{\ensuremath{{\mathcal{#1}}}}

\newcommand{\inmat}[3][C]{\ensuremath{\in{\mathbb{#1}}^{{#2}\times{#3}}}}

\newcommand{\be}{\begin{equation}}
\newcommand{\ee}{\end{equation}}
\newcommand{\bi}{\begin{itemize}}
\newcommand{\ei}{\end{itemize}}
\newcommand{\ma}[1]{\mbox{\boldmath$#1$}}
\newcommand{\ten}[1]{\ma{\mathcal #1}}

\newcommand{\krp}{\diamond}
\newcommand{\kronp}{\otimes}
\newcommand{\beq}{\begin{eqnarray}}
\newcommand{\eeq}{\end{eqnarray}}

\DeclareMathOperator*{\argminA}{arg\,min} 

\title{Tensor-Based Receivers for the Bistatic Sensing and Communication Scenario}
\author{Walter da C. Freitas Jr.$^{1}$, G\'{e}rard Favier$^{2}$ and Andr\'{e} L. F. de
Almeida$^{1}$\\
\textit{\small $^{1}$Laborat\'{o}rio GTEL/DETI/UFC, Campus do Pici, CP 6005, 60455-970, Fortaleza, Cear\'{a}, Brazil}\\
\textit{\small $^{2}$Laboratoire I3S/CNRS/UCA, 2000, Route des Lucioles,
06903 Sophia Antipolis C\'{e}dex, France}\\
\vspace{-2ex} \\ 
\textit{\small E-mails: \{walter,andre\}@gtel.ufc.br, favier@i3s.unice.fr}}

\begin{acronym}[ISAC]
\acro{6G}{Six Generation}
\acro{BS}{base station}
\acro{3GPP}{3rd Generation Partnership Project}
\acro{AoD}{angle-of-departure}
\acro{AoA}{angle-of-arrival}
\acro{DFT}{discrete Fourier transform}
\acro{ISAC}{Integrated sensing and communication}
\acro{UE}{user equipment}
\acro{EVD}{eigenvalue decomposition}
\acro{LS}{least square}
\acro{LOS}{line-of-sight}
\acro{SNR}{signal-to-noise ratio}
\acro{MISO}{multiple-input single-output}
\acro{ULA}{uniform linear array}
\acro{NMSE}{normalized mean square error}
\acro{RIS}{reconfigurable intelligent surfaces}
\acro{KRF}{Khatri-Rao factorization}
\acro{KRST}{Khatri-Rao space-time}
\acro{ALS}{alternating least square}
\acro{CPU}{central process unit}
\acro{XR}{extended reality}
\acro{SER}{symbol error rate}
\acro{SVD}{singular value decomposition}
\acro{MUSIC}{multiple signal classification}
\acro{ZF}{zero-forcing}
\end{acronym}

\begin{document}

\maketitle

\begin{abstract}
We propose receivers for bistatic sensing and communication that exploit a tensor modeling of the received signals. We consider a hybrid scenario where the sensing link knows the transmitted data to estimate the target parameters while the communication link operates semi-blindly in a direct data decoding approach without channel knowledge. We show that the signals received at the sensing receiver and communication receiver follow PARATUCK and PARAFAC tensor models, respectively. These models are exploited to obtain accurate estimates of the target parameters (at the sensing receiver) and the transmitted symbols and channels (at the user equipment). We discuss uniqueness conditions and provide some simulation results to evaluate the performance of the proposed receivers. Our experiments show that the sensing parameters are well estimated at moderate \ac{SNR} while keeping good \ac{SER} {\color{black} and channel \ac{NMSE}} results for the communication link.
\end{abstract}

\section{Introduction}\label{sec1}
\IEEEPARstart{S}{ensing} in current communication networks has been identified as a possible main service for the next generation (6G) wireless systems \cite{3GPP:TS22.837}, \cite{3GPP:TS22.137}. The main advantage of such provision is that the wireless systems dispose of good and diverse infrastructure \cite{Liu:20}. Sensing refers to using radio signals to detect and estimate characteristics of targets in the environment. Integrating sensing into the network nodes means adding a ``radar'' functionality to sense/comprehend the physical world in which they operate. This integration can occur by sharing the hardware and/or waveforms \cite{Emil:book24}. 

{\color{black}
The integration topology depends on the location of the transmitter and receiver, signal awareness of targets, levels of integration (for example, sites, spectrum, hardware, or waveforms), and the entity that transmits the sensing signals, namely, network (NW)-based or user equipment (UE)-based sensing and so on \cite{Xiong:2023}. The traditional radar topology is monostatic, where the same node is used as a transmitter and receiver.
}
We are interested in the bistatic sensing and communication {\color{black}scenario} consisting of two \acp{BS}. The former sends sensing and communication signals; the latter acts as a sensing receiver only. The communication link between the transmitter \ac{BS} and a multiantenna \ac{UE} is established.

{\color{black}
The use of tensor decompositions has been widely studied
for wireless communication systems. The practical motivation for tensor modeling is that one can simultaneously benefit from multiple (more than two) signal diversities, like space, time, and frequency diversities, for instance. Recent works have proposed semi-blind receivers with the joint symbol and channel estimation in different architectures and deployments of wireless communication systems, e.g. cooperative scenario \cite{Leandro:TSP14} \cite{Walter:SPL:17},
massive-MIMO enabled, reconfigurable intelligent surfaces (RIS) nodes \cite{Gilderlan:21}, and so on. The PARAFAC decomposition \cite{Favier:PT2,Favier:bookv2} is the most popular tensor decomposition; however, its simplicity may not capture all tensor relations, which arise in several practical problems. Another popular tensor decomposition sharing properties of the PARAFAC and Tucker decompositions is the so-called PARATUCK  \cite{Harshman:96, Bro:98, Ferrer:09, Gil:23} tensor decomposition. The PARATUCK model is more suitable for complex scenarios where the powerful uniqueness properties of PARAFAC decomposition and the flexibility of the Tucker decomposition are required. Moreover, due to its flexible algebraic structure, it has been efficiently applied to solve problems in signal processing for wireless communication, as shown in \cite{Favier:PT2}. However, to the best of the authors' knowledge, the application of tensor-based semi-blind receivers to the ISAC scenario is still an unexplored topic. 

In \cite{Zhang:24}, the authors consider a massive MIMO monostatic sensing scenario representing the collecting of echo signals as a third-order tensor following a PARAFAC model to estimate the environment parameters, including
angles, time delays, Doppler shifts, and reflection/path coefficients
of the targets/channels. The work \cite{Du:24} evaluates an ISAC scenario with a \ac{RIS}-assisted downlink terahertz (THz) multiple-input multiple-output (MIMO) system. The authors formulate the received signal at the vehicle terminal as a nested tensor that is composed of multiple outer parallel factors (PARAFAC) tensors and an inner PARATUCK tensor.
}

In this work, we show that the signals received at the sensing and communication receivers follow PARATUCK and PARAFAC tensor models, respectively, from which the estimation of the target parameters, communication channel, and transmitted data symbols are obtained. For the sensing part, an iterative receiver is proposed to jointly estimate the target parameters (\ac{AoA}, \ac{AoD} and reflection coefficients) by exploiting a PARATUCK2 tensor model.
A closed-form semi-blind receiver using rank-one matrix approximations is derived based on a PARAFAC tensor model for the communication link. We summarize the identifiability conditions and uniqueness issues for the proposed estimation methods. Our numerical results show that the sensing parameters are well estimated at moderate \ac{SNR} 
with good \ac{SER} and \ac{NMSE} results for the communication link operating semi-blindly.

{\color{black}

\textit{\textit{Notation}}: Scalars, column vectors, matrices, and tensors are denoted by lower-case, boldface lower-case, boldface
upper-case, and calligraphic letters, e.g., $a, \mat a, \mat A, \matc A$, respectively. $\mat A_{i.}$ and $\mat A_{.j}$ represent the $i$-th row and the $j$-th column of $\mat A\inmat{I}{J}$, respectively. The operator $vec(·)$ transforms a matrix into a column vector by stacking the columns of its matrix argument, $D_n(\mat A)$ is a diagonal matrix with diagonal entries given by $n-$th row of $\mat A$. The {\color{black} Khatri-Rao and Kronecker products are denoted by $\krp$} and $\otimes$, respectively.
The identity and all-zeros matrices of dimensions $N \times N$ are denoted as $\mat I_{N}$ and $\mat 0_{N}$, respectively.
We use the superscripts $^T, ^*, ^H, ^{-1}, ^{\dagger}$ for matrix transposition, complex conjugation, Hermitian transposition, inversion, and Moore-Penrose pseudo inversion, respectively. }%

\section{System model and assumptions}
\label{sec:chan_model}
\begin{figure}[!t]
    \centering
    \includegraphics[width=7cm]{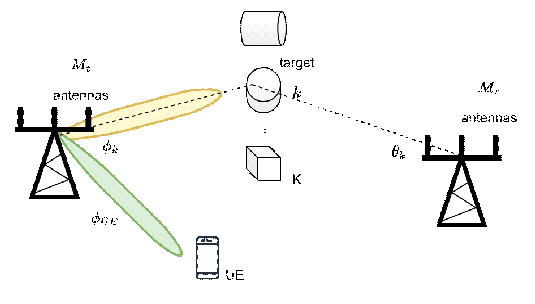}
    \caption{Bistatic sensing and communication scenario.}\label{Fig:scenario}
\end{figure}

We consider a bistatic sensing and communication system composed
of one \ac{BS} as transmitter with $M_t$ transmit antennas, and a second as a sensing receiver with $M_r$ receive antennas, where \ac{LOS} between the two BSs is assumed {\color{black} to be unavailable}, as well as $K$ targets, see Figure~\ref{Fig:scenario}. 

 Let, $\mat{A}_{R}(\Theta)=\left[\mat{a}_{R}(\theta_1)\quad \cdots \quad \mat{a}_{R}(\theta_K)\right]\in \mathbb{C}^{M_r \times K}$ and $\mat{A}_{T}(\Phi)=\left[\mat{a}_{T}(\phi_1)\quad \cdots \quad \mat{a}_{T}(\phi_K)\right]\in \mathbb{C}^{M_t \times K}$ be the receiver and transmit steering matrices {\color{black}for the channel between the BSs, via the targets}, respectively. The matrix $\mat{\Gamma} \in \mathbb{C}^{N \times K}$ contains the reflection coefficients of the $K$ targets, such that in the $n-$th time-slot for the $k-$th target one have $\gamma_{n,k} \sim \mathcal{CN}(0,\sigma^2)$. {\color{black}
  The transmitted pilot symbols are $\mat S^{(p)} \in \mathbb{C}^{P \times M_t}$ while the data symbols are $\mat S \in \mathbb{C}^{P \times M_t}$} and both are precoded by the \ac{KRST} code matrix $\mat C \in \mathbb{C}^{N \times M_t}$. The signals received at sensing \ac{BS} 
define a third-order tensor $\mathcal{Y} \in \mathbb{C}^{M_{r} \times P \times N}$ following a PARATUCK-2 decomposition \cite{Favier:bookv2} with $\mat{A}_{R}$ and $\mat C$ as matrix factors, $\mat{A}^T_{T}$ as core matrix and $\mat \Gamma$ and $\mat S$ as interaction matrices, such that, the $n-$th frontal slice of $\cal Y$ is given by 
\begin{equation}
\mat{Y}_{..n} = \mat{A}_{R}(\Theta)D_n(\mat \Gamma)\mat{A}^T_{T}(\Phi) D_n(\mat C) \mat S{^{(p)}}^T \in \mathbb{C}^{M_r \times P}\label{fslice:TenYn}.
\end{equation}
Assuming the \ac{UE} with $M_u$ receive antennas, the signals received at \ac{UE}, at time slot $n$,  
define the $n-$th frontal slice of the received signal tensor $\mathcal{Y}^{(UE)}\in \mathbb{C}^{M_u \times P \times N}$ given by{\color{black}
\begin{equation}
\mat{Y}^{(UE)}_{..n} = \mat{H} D_{n}(\mat C) \mat S^T \in \mathbb{C}^{M_u \times P}, 
\end{equation}
where $\mat{H}=\mat{A}_{R}(\Theta_{UE})\mat G\mat{A}^T_{T}(\Phi_{UE})\in \mathbb{C}^{M_{u} \times M_t}$ is the effective communication channel and {\color{black}$\mat G=diag(g_1,\cdots,g_L)\in \mathbb{C}^{L \times L}$} is a diagonal matrix that contains the $L$ path gains.}

\section{Tensor-based receivers}
{\color{black}
The transmitted pilot symbols $\mat S{^{(p)}}$ and the precoded \ac{KRST} code matrix $\mat C$ are assumed to be known to the sensing receiver, while for the communication receiver, only the \ac{KRST} code matrix $\mat C$ is known.}

Defining $\mat F_{..n}=D_n(\mat \Gamma)\mat{A}^T_{T}(\Phi) D_n(\mat C)\mat S{^{(p)}}^T \in \mathbb{C}^{K \times P}$ and $\mat F = \left[\mat F_{..1}\quad \mat F_{..2} \quad \cdots \quad \mat F_{..N} \right] \in \mathbb{C}^{K \times NP}$. A flat 1-mode unfolding of $\mathcal{Y}$ is obtained from Eq.~(1) as: $\mat Y_{(1)} = \left[\mat Y_{..1}\quad \mat Y_{..2} \quad \cdots \quad \mat Y_{..N} \right] = \mat{A}_{R}(\Theta)\mat F \in \mathbb{C}^{M_r \times NP}$, and a \ac{LS} estimate of the steering matrix $\mat{A}_{R}(\Theta)$ is given by: 
\begin{equation}\label{Eq.Ar_est}
\hat{\mat{A}}_{R}(\Theta) = \mat{Y}_{(1)} \mat F^{\dag}.
\end{equation}
Applying the $vec(\cdot)$ operation to Eq.~(\ref{fslice:TenYn}) gives:
$
vec\left(\mat{Y}_{..n}\right) =\big[(\mat S{^{(p)}} D_n(\mat C)) \kronp (\mat{A}_{R}(\Theta)D_n(\mat \Gamma))\big]vec(\mat{A}^T_{T}(\Phi))\in \mathbb{C}^{PM_{r}}$. Stacking row-wise these vectors for $N$ time slots leads to
\begin{equation}
\underbrace{\left[
\begin{matrix}
vec(\mat{Y}_{..1})\\
vec(\mat{Y}_{..2})\\
\vdots\\
vec(\mat{Y}_{..N})
\end{matrix}
\right]}_{\mat y} =
\underbrace{\left[
\begin{matrix}
\mat S{^{(p)}} D_1(\mat C) \kronp \mat{A}_{R}(\Theta)D_1(\mat \Gamma)\\
\vdots\\
\mat S{^{(p)}} D_N(\mat C) \kronp \mat{A}_{R}(\Theta)D_N(\mat \Gamma)
\end{matrix}
\right]}_{\mat M} vec(\mat{A}^T_{T}(\Phi)).
\end{equation}
{\color{black}
Therefore, 
\begin{equation}\label{Eq.At_est}
 vec(\hat{\mat{A}}^T_{T}(\Phi)) = \mat M^{\dag}\mat y.
\end{equation}
}

Defining $\mat G_{..n} = \mat S{^{(p)}}  D_n(\mat C) \mat{A}_{T}(\Phi) \in \mathbb{C}^{P \times K}$, yields $vec\left(\mat{Y}_{..n}\right) = \left[\mat G_{..n} \krp \mat{A}_{R}(\Theta)\right]\mat \Gamma_{n.}^T$. {\color{black} The LS estimate of} $\mat \Gamma_{n.}$ is obtained as 
\begin{equation}\label{Eq.Gamma}
\hat{\mat\Gamma}_{n.} = \left[\left(\mat G_{..n} \krp \mat{A}_{R}(\Theta)\right)^{\dag}vec\left(\mat{Y}_{..n}\right)\right]^T, \,\, n=1,\ldots, N.
\end{equation}
{\color{black}
A three-step \ac{ALS}-based receiver \cite{Comon2009} is applied to estimate $\mat{A}_{R}(\Theta)$, $\mat{A}_{T}(\Phi)$, and $\mat \Gamma$ iteratively. The steering matrices and reflection coefficients are jointly estimated by alternately
minimizing the following LS criteria
\begin{eqnarray}
\hat{\mat{A}}_{R}(\Theta)&=&\argminA_{\mat{A}_{R}}\|
\mat Y_{(1)}- \mat{A}_{R}\mat F\|^{2}_{F},\\
\hat{\mat{A}}_{T}(\Phi)&=& \argminA_{\mat{A}_{T}}\|\mat y - \mat Mvec(\mat{A}_{T}^T)\|^{2}_{2},\\
\hat{\mat{\Gamma}}_{n.}&=& \argminA_{\mat\Gamma_{n.}}\|  vec\left(\mat{Y}_{..n}\right) - \mat{\Gamma}_{n.}^T\left(\mat G_{..n} \krp \mat{A}_{R}\right)\|^{2}_{2}.
\end{eqnarray}
The process is repeated until convergence is reached.
}
{\color{black}
The tensor $\mathcal{Y}^{(UE)}$ defining the received signal at the \ac{UE} satisfies a PARAFAC model with the following factors $[[\mathbf{H},\mat S,\mathbf{C};M_t]]$. A tall 3-mode unfolding of  $\mathcal{Y}^{(UE)}$ is given by: $\mat Y_{PM_u \times N}^{(UE)} =\left(\mat S\krp \mat{H} \right)\mat C^T$ from which we deduce the following
\ac{LS} estimate:
\begin{equation}\label{Eq.Q}
    \hat{\mat{Q}}=\mat Y_{PM_u \times N}^{(UE)} \mat{C}^* \,\approx \mat S\krp \mat{H},
\end{equation}
{\color{black}assuming the coding matrix $\mat{C}$ is designed as column orthonormal matrix.}

The UE can jointly recover the transmitted symbol matrix $\mat S$ and the channel matrix $\mathbf{H}$ in a closed-form by solving the so-called \ac{KRF} problem \cite{Favier:bookv2,Kibangou:20109}
$(\hat{\mat{S}},\hat{\mat{H}})=\underset{\mathbf{S},\mathbf{H}}{\textrm{argmin}}=\|\hat{\mat{Q}} - \mat S\krp \mat{H}\|^2_F.$
The tensor-based receivers for the bistatic sensing and communication scenario are summarized in Table~\ref{Tab_Rx_K2}.

}
{\color{black}
\begin{table}
\caption{Tensor-based bistatic sensing and communication receiver.}
\label{Tab_Rx_K2}
\noindent\rule{.98\columnwidth}{.1pt}
\begin{enumerate}
\item \textbf{Sensing parameters estimation}: a priori information -- received tensor $\ten{Y}$, pilot symbol matrix $\mat S^{(p)}$ and code matrix $\mat{C}$.
\begin{itemize}
\item[(1.1)] Set $i = 0$ and initialize randomly $\mat{A}_{R}(\Theta)$, $\mat{\Gamma}$ and $\mat{A}_{T}(\Phi)$;
\item[(1.2)] $i\longleftarrow i + 1$;
\item[(1.3)] From 1-mode unfolding of $\ten{Y}$, calculate the matrices $\mat Y$ and $\mat F$ to obtain an LS estimate of $\mat{A}_{R}(\Theta)$ from (\ref{Eq.Ar_est}) as:
\begin{equation*}
 \hat{\mat{A}}_{R(i)} = \mat{Y}_{(1)_{}} \left(\mat F_{(i-1)}\right)^{\dag}
\end{equation*}
\item[(1.4)] Stacking row-wise $vec\left(\mat{Y}_{..n}\right)$ for the $N$ time-slots to calculate $\mat M$, we obtain an LS estimate of $\mat{A}_{T}(\Phi)$ from (\ref{Eq.At_est}) as:
\begin{equation*}
 vec(\hat{\mat{A}}^T_{T_{(i)}}) = \mat M_{}^{\dag}\mat y.
\end{equation*}
\item[(1.5)] Calculate $\mat G_{..n}$ to obtain a LS estimate of $\mat \Gamma_{n.}$ ($n=1,\ldots, N$) from (\ref{Eq.Gamma}) as 
\begin{equation*}\label{Eq.Gamma}
\hat{\mat\Gamma}_{n._{(i)}} = \left[\left(\mat G_{..n_{(i)}} \krp \mat{A}_{R_{(i)}}(\Theta)\right)^{\dag}vec\left(\mat{Y}_{..n}\right)\right]^T.
\end{equation*}
\item[(1.6)] Repeat steps 1.3-1.5 until convergence.
\end{itemize}
\item \textbf{Symbol communication estimation}: 
\begin{itemize}
\item[(2.1)] Compute the LS estimate $\hat{\mat{Q}}=\mat Y_{PM_r \times N}^{(UE)} \mat{C}^*$.
\item[(2.2)] Apply the KRF algorithm to estimate  $\mat S$ and $\mat{H}$ from $\hat{\mat{Q}}$.
\item[(2.3)] Remove the scaling ambiguities from $\mat S$ and $\mat{H}$, respectively.
\end{itemize}
\end{enumerate}
\noindent\rule{.98\columnwidth}{.1pt}
\end{table}
}

\subsection{Identifiability and uniqueness}
{\color{black}
The identifiability and uniqueness conditions of PARATUCK2 decomposition can be found in 
\cite{Favier:PT2}.}
From Eqs. (\ref{Eq.Ar_est}) and (\ref{Eq.At_est}), we find that a unique estimation of the AoA and AoD parameters require $NP \geq K$ and $NPM_r\geq M_tK$. Combining both inequalities, we conclude that the maximum number $K$ of targets supported by the proposed receiver is given by $K_{\textrm{max}}=\textrm{min}(NP,NPM_r/M_t)$. {\color{black} From (\ref{Eq.Gamma}) the uniqueness is obtained if $PM_r \geq K$.} For the communication link, the first estimation step in (\ref{Eq.Q}) requires $N\geq M_t$. The uniqueness of the sensing signal model (\ref{fslice:TenYn}) is ensured up to column scaling and permutation ambiguities deﬁned by the following relations:
\begin{eqnarray}
\bar{\mat{A}}_{R}(\Theta) &=& \mat{A}_{R}(\Theta)\left(\mat P \mat \Lambda^{(\mat{A}_{R})} \right),\\
D_n(\bar{\mat{\Gamma}}) &=&\left(z_n ^{-1} \mat P^{T}\right)D_n(\mat \Gamma)\left(\mat P \mat \Lambda^{(\mat \Gamma)} \right),\\
\bar{\mat{A}}^T_{T}(\Phi) &=& \left(\mat{\Lambda}^{(\mat \Gamma)}\right)^{-1}\left(\mat{\Lambda}^{(\mat{A}_{R})}\right)^{-1}\mat P^T\mat{A}^T_{T}(\Phi),
\end{eqnarray}
where $\mat{\Lambda}^{(\mat{A}_{R})}$, $\mat{\Lambda}^{(\mat \Gamma)}$ are (diagonal) scaling matrices, $\{z_n\}$, $n=1,\ldots, N$, are scalars, and $\mat P\in \mathbb{C}^{K \times K}$ is a permutation matrix. 
The knowledge of $\mat C$ and $\mat S^{(p)}$ (pilot symbols for sensing) implies $\mat{\Lambda}^{(\mat{A}_{R})}=\textrm{diag}^{-1}(\hat{\mat{A}}_{{R}_{1.}})$, $\mat{\Lambda}^{(\mat \Gamma)} = (\mat{\Lambda}^{(\mat{A}_{R})})^{-1}$. The permutation matrix does not represent a problem for the angle estimation. The symbol and channel estimates obtained at the UE are affected by scaling ambiguities, which can be compensated by the knowledge of the first row of matrix $\mat S$.

{\color{black}
\subsection{Complexity analysis}
The dominant complexity of the proposed receivers is associated with the computation of the matrix pseudo-inverses in three \ac{LS} update steps that calculate the estimates of the steering matrices in an iterative and alternating way. 

Note that, for a matrix of dimensions $J \times K$, the complexity of its \ac{SVD} computation is $\mathbb{O}(\min(J,K)JK)$. Hence, the complexity of the proposed receiver is dominated by steps (1.3) and (1.4). The steps (1.3) and (1.4) have complexity $\mathbb{O}(\min(K,NP)KNP)$ and $\mathbb{O}(\min(PM_r,N)PM_rN)$, respectively.
}

\section{Numerical results}
We assume that the numbers of transmit and receive (sensing and communication) antennas are equal to $M_t=M_r= 2$, the number of targets is $K = 2$, the total number of sensing and communication slots is $N = 3$, and $P=8$. The \acp{AoD} and \acp{AoA} are centered at $-37^{\circ}$ and $65^{\circ}$, and $15^{\circ}$ and $27^{\circ}$, respectively. The \ac{AoD} and \ac{AoA} for the communication link are set to $25^{\circ}$ and $78^{\circ}$, respectively. The transmitted symbols are randomly drawn from a $4$-QAM alphabet, and for the sensing link, they are considered known at the \ac{BS} receiver for sensing purposes. At each Monte Carlo run, the  $E_s/N_0$ ratio is controlled by fixing $E_s=1$ and varying $N_0$ to ensure the desired $E_s/N_0$ value.

{\color{black}

Figure~2 presents the steering vectors' estimations for the \ac{AoA} target parameters in different values of \ac{SNR}. At a \ac{SNR}s of 20dB, we achieve remarkable estimates of the steering vectors. Similar results are achieved for the \ac{AoD} estimation.

In Figure~3, we evaluate the performance of steering vectors' estimation, now considering different numbers of time slots and assuming an \ac{SNR} of 10dB. We can see the impact of the number of time slots on the estimation accuracy. With $N=16$, we obtain very accurate estimates for an \ac{SNR} around 10dB.

In Figure~4, we still evaluate the steering vectors' estimations but now consider different multiples of pilot symbols ($P=4$) while assuming an \ac{SNR} of 10dB. The result shows the impact of the number of pilot symbols on the estimation accuracy. For instance, with $P=16$ pilot symbols, a very good result is obtained.

Figure~5 depicts the convergence behavior of the proposed \ac{ALS} algorithm as a function of the iterations for different \ac{SNR}s. We can note that, even for a low \ac{SNR}, the algorithm converges with a low \ac{NMSE}. As a stopping criterion, we declare the convergence when the difference between the reconstruction error in two successive iterations is smaller than $10^{-6}$. A maximum number of 1000 iterations is assumed.
Figure~6 presents the \acp{NMSE} of the estimated angles. As a comparison, we include the performance of the angles' estimates using the rectification method from \cite{Couras:2023}. This rectification method consists of constructing a rank-one Hermitian Toeplitz matrix from each column of the estimated matrix and computing its \ac{EVD}. The angles reconstructed using the method of \cite{Couras:2023} present a small improvement in the low \ac{SNR} regime, i.e., below 15dB.

Figure~7 evaluates the communication performance in terms of symbol error rate (SER). The transmitted symbols are randomly drawn from a $4$-QAM alphabet, assuming $M_u=2$. As a benchmark, we also consider the 
optimal \ac{ZF} beamformer, which assumes perfect knowledge of the steering vectors. As expected, the SER is improved when all channel directions are known, which corresponds to perfect beam steering. We can also see the impact of the number of adding more receive antennas on the communication performance, as expected. Finally, Figure~8 evaluates the channel estimation performance for the communication link in terms of the \ac{NMSE} of the downlink BS-\ac{UE} channel for different numbers of antennas at the \ac{UE}. At moderate \ac{SNR}, accurate channel estimation is obtained, corroborating the good \ac{SER} performance for the communication link.
}

\begin{figure}[!t]
\centering
\includegraphics[width=6.cm]{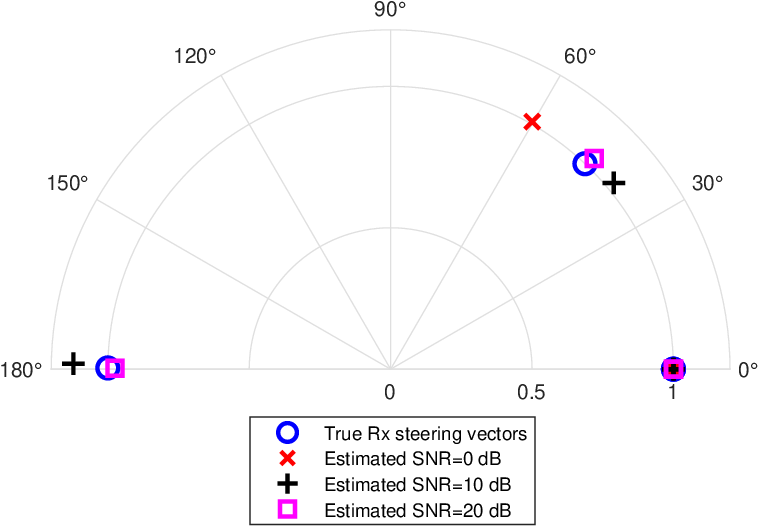}
    \caption{Performance of AoA steering vectors’ estimation with different \ac{SNR}s.}
\end{figure}

\begin{figure}[!t]
    \centering
    \includegraphics[width=6.cm]{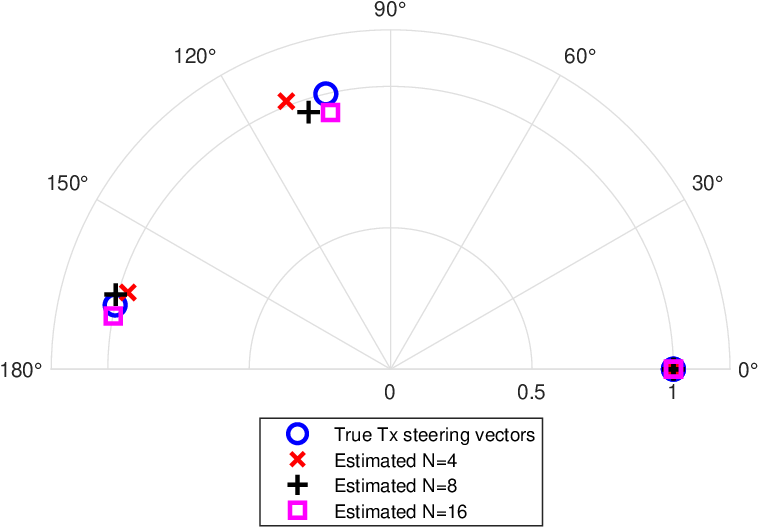}
   \caption{Performance of AoD steering vectors’ estimation with different time-slots (SNR=10 dB).}
\end{figure}

\begin{figure}[!t]
\centering
    \begin{subfigure}[b]{0.45\textwidth}
    \centering
    \includegraphics[width=6.cm]{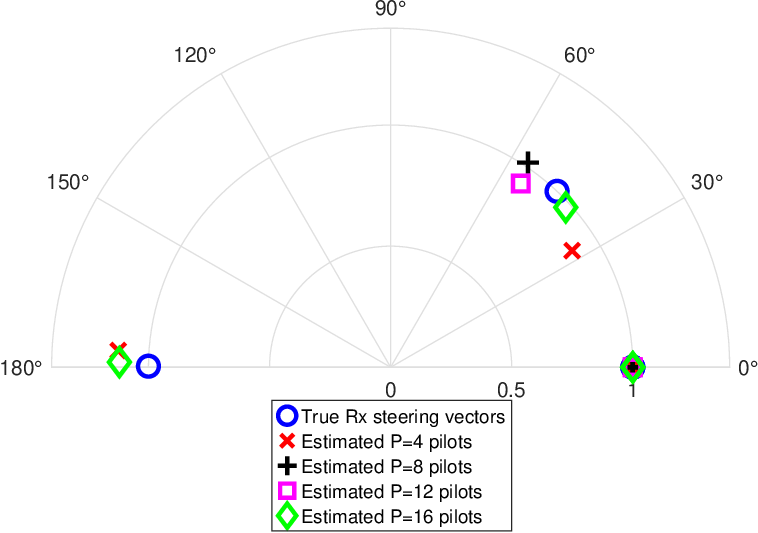}
    \caption{Performance of AoA steering vectors’ estimation with different pilot symbols.}
    \end{subfigure}
\quad
    \begin{subfigure}[b]{0.45\textwidth}
    \centering
    \includegraphics[width=6.cm]{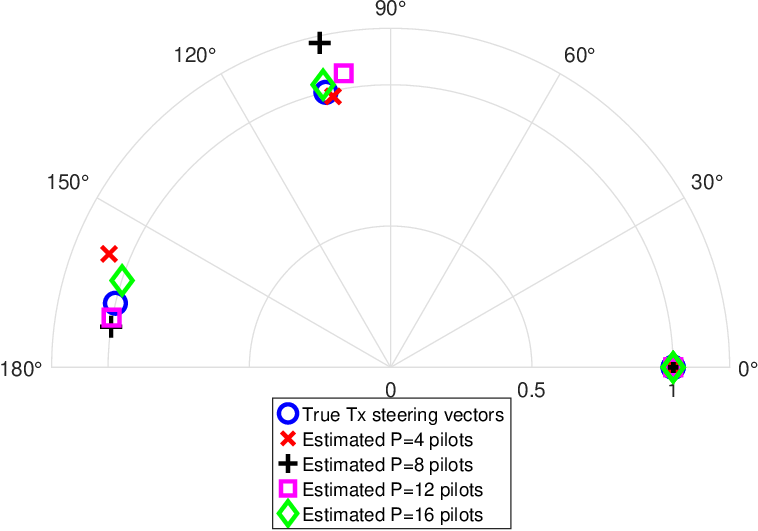}
   \caption{Performance of AoD steering vectors’ estimation with different pilot symbols.}
    \end{subfigure}
    \caption{Performance of steering vectors’ estimation for different numbers of pilot symbols at SNR=10 dB.}
\end{figure}

\begin{figure}[!t]
    \centering
    \includegraphics[width=6.5cm]{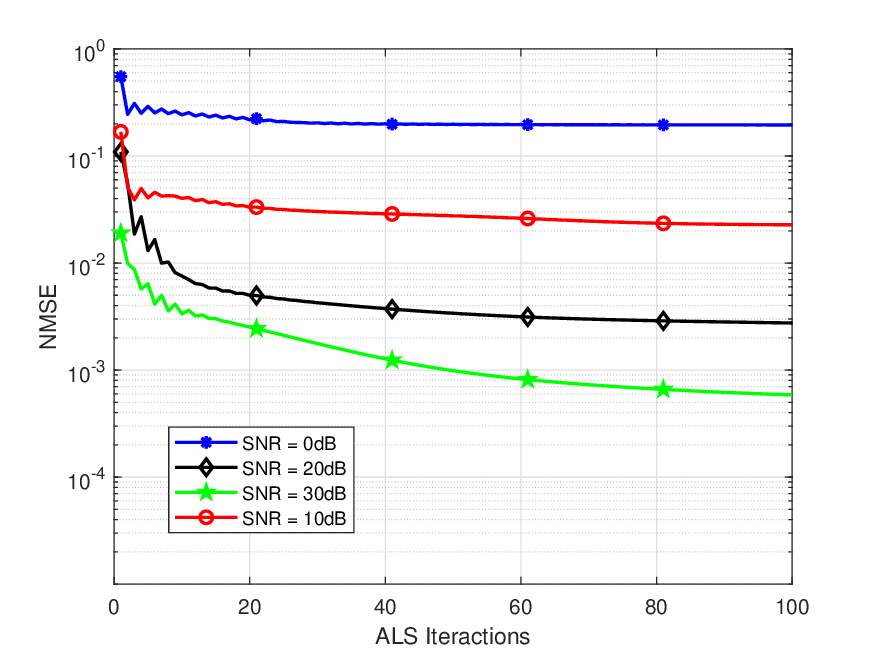}
    \caption{NMSE performance for the reconstructed tensor as a function of ALS iterations.}
\end{figure}

\begin{figure}[!t]
    \centering
    \includegraphics[width=6.5cm]{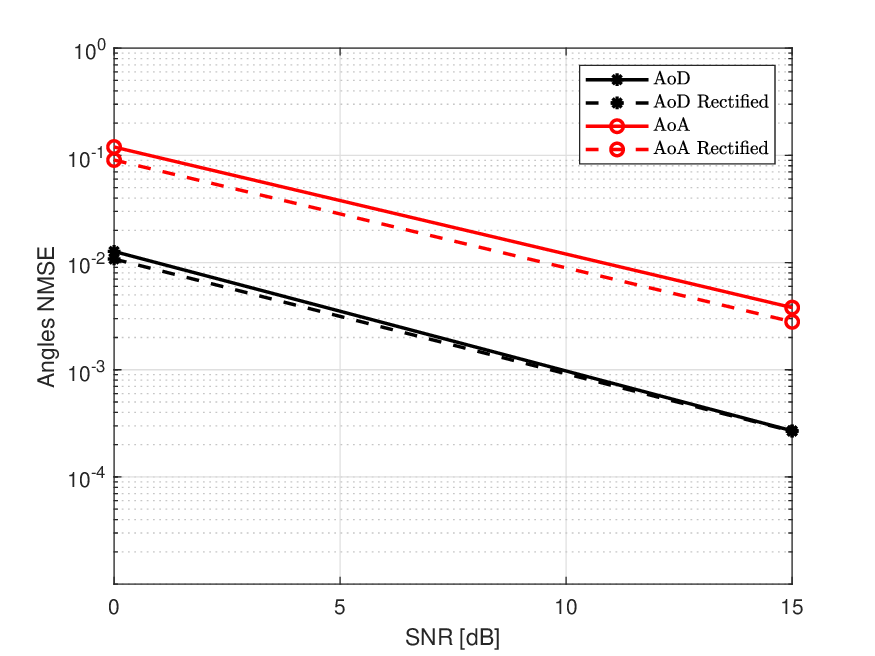}
    \caption{NMSE performance for the estimated angles.}
\end{figure}

\begin{figure}[!t]
    \centering
    \includegraphics[width=6.5cm]{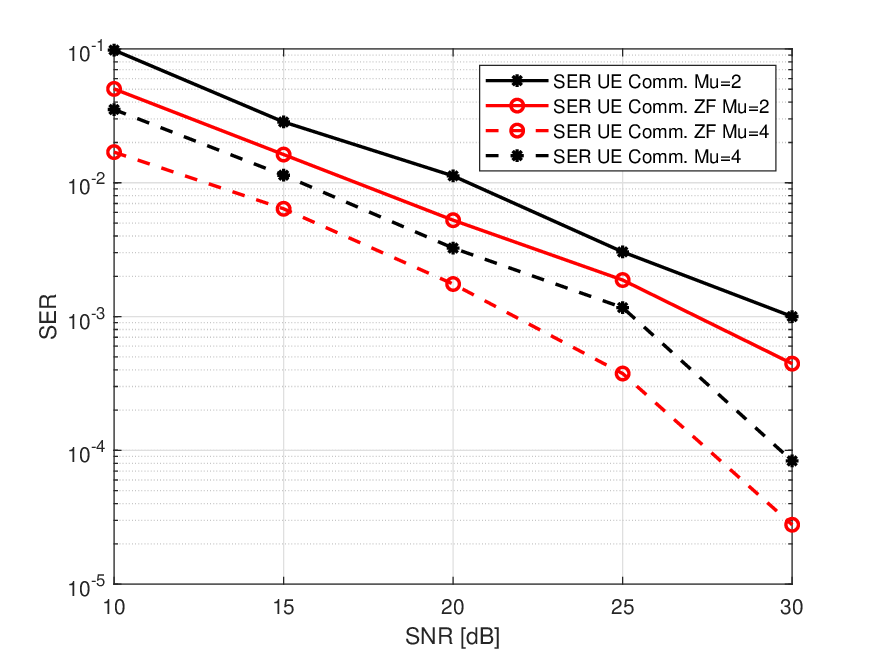}
    \caption{SER performance for the communication link for different numbers of receiver antennas.}
\end{figure}

\begin{figure}[!t]
    \centering
    \includegraphics[width=6.5cm]{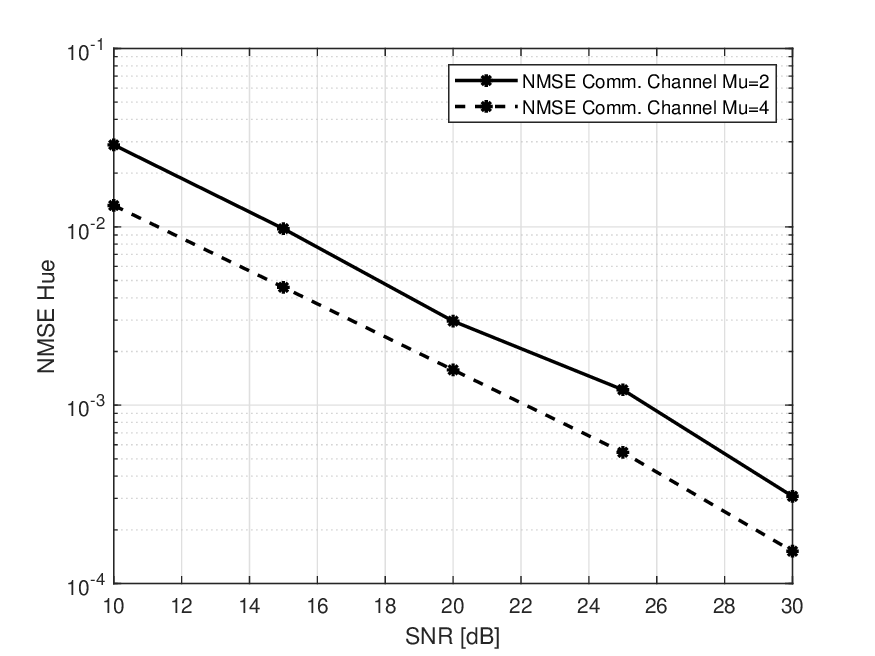}
    \caption{NMSE performance for the communication channel for different numbers of receiver antennas.}
\end{figure}

\section{Conclusions}
We have presented tensor-based receivers for the bistatic sensing and communication scenario. We have addressed a hybrid scenario where the sensing link knows the transmitted data to estimate the target parameters while the communication link operates semi-blindly in a direct data decoding approach without channel knowledge. By exploiting PARATUCK and PARAFAC tensor models for the sensing and communication links, respectively, the proposed tensor-based receiver algorithms estimate the target parameters, communication channel, and transmitted data symbols.
Simulation results have illustrated the proposed receivers' performance in estimating the target parameters using actual information symbols while providing accurate channel estimation and symbol detection for the communication link. 



\end{document}